\documentclass[a4paper,twoside]{article}

\usepackage{epsfig}
\usepackage{subfigure}
\usepackage{calc}
\usepackage{amssymb}
\usepackage{amstext}
\usepackage{amsmath}
\usepackage{amsthm}
\usepackage{multicol}
\usepackage{multirow, makecell}
 \usepackage{multirow}
 \usepackage{xcolor}
 \usepackage{booktabs}
 \usepackage{float}
 \usepackage{subfigure}
 \usepackage{acronym}
 \usepackage{url}
 \usepackage{comment}
\usepackage{pslatex}
\usepackage{apalike}
\usepackage{comment}
\usepackage{SCITEPRESS}  
\usepackage{fancyhdr}
\begin{document}
\title{Interactive Lungs Auscultation with Reinforcement Learning Agent}

\author{\authorname{Tomasz Grzywalski\sup{1}, Riccardo Belluzzo\sup{1}, Szymon Drgas \sup{1}\sup{2}, Agnieszka Cwali\'nska\sup{3} and Honorata Hafke-Dys\sup{1}\sup{4}}
\affiliation{\sup{1}StethoMe\textsuperscript{\textregistered} Winogrady 18a, 61-663 Pozna\'n, Poland}
\affiliation{\sup{2}Institute of Automation and Robotics, Pozna\'n University of Technology, Piotrowo 3a 60-965 Pozna\'n, Poland}
\affiliation{\sup{3}PhD Department of  Infectious Diseases and Child Neurology Karol Marcinkowski, University of Medical Sciences}
\affiliation{\sup{4}Institute of Acoustics, Faculty of Physics, Adam Mickiewicz University, Umultowska 85, 61-614 Pozna\'n, Poland}
\email{\{grzywalski, belluzzo, drgas, hafke\}@stethome.com, szymon.drgas@put.poznan.pl, agnieszka.cwalinska@kmu.poznan.pl, h.hafke@amu.edu.pl}
}

\abstract{To perform a precise auscultation for the purposes of examination of respiratory system normally requires the presence of an experienced doctor. With most recent advances in machine learning and artificial intelligence, automatic detection of pathological breath phenomena in sounds recorded with stethoscope becomes a reality. But to perform a full auscultation in home environment by layman is another matter, especially if the patient is a child. In this paper we propose a unique application of Reinforcement Learning for training an agent that interactively guides the end user throughout the auscultation procedure. We show that \textit{intelligent} selection of auscultation points by the agent reduces time of the examination fourfold without significant decrease in diagnosis accuracy compared to \textit{exhaustive} auscultation.}

\onecolumn \maketitle \normalsize \vfill

\section{\uppercase{Introduction}}
\label{sec:introduction}

Lung sounds auscultation is the first and most common examination carried out by every general practitioner or family doctor.  It is fast, easy and well known procedure, popularized by La\"ennec \cite{stethoscopeInventor}, who invented the stethoscope. Nowadays, different variants of such tool can be found on the market, both analog and electronic, but regardless of the type of stethoscope, this process still is highly subjective. Indeed, an auscultation normally involves the usage of a stethoscope by a physician, thus relying on the examiner's own hearing, experience and ability to interpret psychoacoustical features. Another strong limitation of standard auscultation can be found in the stethoscope itself, since its frequency response tends to attenuate frequency components of the lung sound signal above nearly $120 \ Hz$, leaving lower frequency bands to be analyzed and to which the human ear is not really sensitive \cite{intro1} \cite{auscultation}. A way to overcome this limitation and inherent subjectivity of the diagnosis of diseases and lung disorders is by digital recording and subsequent computerized analysis \cite{AI4lungs}.

Historically many efforts have been reported in literature to automatically detect lung sound pathologies by means of digital signal processing and simple time-frequency analysis \cite{AI4lungs}. In recent years, however, machine learning techniques have gained popularity in this field because of their potential to find significant diagnostic information relying on statistical distribution of data itself \cite{intro2}. Palaniappan et al. (2013) report state of the art results are obtained by using supervised learning algorithms such as support vector machine (SVM), decision trees and artificial neural networks (ANNs) trained with expert-engineered features extracted from audio signals. However, more recent studies \cite{CNNsLungSounds} have proved that such benchmark results can be obtained through end-to-end learning, by means of deep neural networks (DNNs), a type of machine learning algorithm that attempts to model high-level abstractions in complex data, composing its processing from multiple non-linear transformations, thus incorporating the feature extraction itself in the training process. Among the most successful deep neural network architectures, convolutional neural networks (CNNs) together with recurrent neural networks (RNNs) have been shown to be able to find useful features in the lung sound signals as well as to track temporal correlations between repeating patterns \cite{CNNsLungSounds}. 

However, information fusion between different auscultation points (APs) and integration of decision making processes to guide the examiner throughout the auscultation seems to be absent in the literature. Reinforcement Learning (RL), a branch of machine learning inspired by behavioral psychology \cite{Shteingart2014ReinforcementLA}, can possibly provide a way to integrate auscultation path information, interactively, at data acquisition stage. In the common RL problem setting, the algorithm, also referred as agent, learns to solve complex problems by interacting with an environment, which in turn provides positive or negative rewards depending on the results of the actions taken. The objective of the agent is thus to find the best \textit{policy}, which is the best action to take, given a state, in order to maximize received reward and minimize received penalty. We believe that RL framework is the best choice for the solution of our problem, i.e finding the lowest number of APs while maintaining a minimum acceptable diagnosis accuracy. As a result, the agent will learn what are the optimal locations to auscultate the patient and in which order to examine them. As far as we know, this is the first attempt to use RL to perform interactive lung auscultation. \\
\indent RL has been successful in solving a variety of complex tasks, such as computer vision \cite{RLAcomputervision}, video games \cite{RLAvideogames}, speech recognition \cite{RLAspeech} and many others. RL can also be effective in feature selection, defined as the problem of identifying the smallest subset of highly predictive features out of a possibly large set of candidate features \cite{HAZRATIFARD20131892}. A similar problem was further investigated \cite{7175757} where the authors develop a Markov decision process (MDP) \cite{Puterman:1994:MDP:528623} that through dynamic programming (DP) finds the optimal feature selecting sequence for a general classification task. Their work motivated us to take advantage of this framework with the aim of applying it to find the lowest number of APs, i.e the smallest set of features, while maximizing the accuracy in classifying seriousness of breath phenomena detected during auscultation, which is in turn directly proportional to diagnosis accuracy.

This work is organized as follows. Section \ref{sec:mathematics} recalls the mathematical background useful to follow the work. Section \ref{sec:RL-based-interactive-auscultation} formally defines our proposed solution and gives a systematic overview of the interactive auscultation application. Section \ref{evaluation} describes the experimental framework used to design the interactive agent and evaluate its performance. Section \ref{results} shows the results of our experiments, where we compare the interactive agent against its static counterpart, i.e an agent that always takes advantage of all auscultation points. Finally, Section \ref{conclusions} presents our conclusions.

\section{\uppercase{Mathematical Background}}
\label{sec:mathematics}

\subsection{Reinforcement Learning}
\label{rl_theory}
\begin{figure}[ht!]
	\centering
		\includegraphics[scale=0.95]{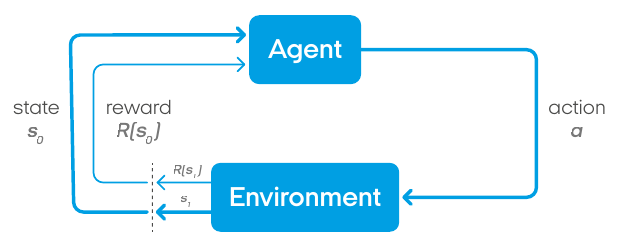}
	\caption{RL general workflow: the agent is at state $s_0$, with reward $R[s_0]$. It performs an action $a$ and goes from state $s_0$ to $s_1$ getting the new reward $R[s_1]$.}
	\label{fig:RL-workflow}
\end{figure}

The RL problem, originally formulated in \cite{Sutton1988}, relies on the theoretical framework of MDP, which consists on a tuple of $(S, A, P_{SA}, \gamma, R)$ that satisfies the Markov property \cite{Puterman:1994:MDP:528623}. $S$ is a set of environment states, $A$ a set of actions, $P_{SA}$ the state (given an action) transitions probability matrix, $\gamma$ the discount factor, $R(s)$ the reward (or reinforcement) of being in state $s$. We define the policy $\pi(s)$ as the function, either deterministic or stochastic, which dictates what action to take given a particular state. We also define a \textit{value function} that determines the value of being in a state $s$ and following the policy $\pi$ till the end of one iteration, i.e an episode. This can be expressed by the expected sum of discounted rewards, as follows:
\begin{equation}
    V^{\pi}(s) = E[R(s_0) + \gamma R(s_1) + \gamma^2 R(s_2) + ... | s_0 = s, \pi ]
\end{equation}
\noindent where $s_0, \ s_1, \ s_2, \ \dots$ is a sequence of states within the episode. The discount factor $\gamma$ is necessary to moderate the effect of observing the next state. When $\gamma$ is close to $0$, there is a shortsighted condition; when it tends to $1$ it exhibits farsighted behaviour \cite{Sutton:1998:IRL:551283}. For finite MDPs, policies can be partially ordered, i.e $\pi\geq\pi'$ if and only if $V^\pi (s) \geq V^{\pi'} (s)$ for all $s \in S$. There is always at least one policy that is better than or equal to all the others. This is called optimal policy and it is denoted by $\pi^*$. The optimal policy leads to the optimal value function:
\begin{align}
 V^*(s) = \max_\pi V^\pi (s) = V^{\pi^*}(s)
\end{align}
\indent In the literature the algorithm solving the RL problem (i.e finding $\pi^*$)  is normally referred as the agent, while the set of actions and states are abstracted as belonging to the environment, which interacts with the agent signaling positive or negative rewards (Figure \ref{fig:RL-workflow}). Popular algorithms for its resolution in the case of finite state-space are \textit{Value iteration} and \textit{Policy iteration} \cite{Sutton1988}.

\subsection{Q-Learning}
\label{q-learn}
Q-learning \cite{Watkins92q-learning}  is a popular algorithm used to solve the RL problem. In Q-learning actions $a \in A$ are obtained from every state $s \in S$ based on an action-value function called \textit{Q function}, $Q: S \times A \rightarrow{\mathbb{R}}$, which evaluates the quality of the pair $(s,a)$.

The Q-learning algorithm starts arbitrarily initializing $Q(s,a)$; then, for each episode, the initial state $s$ is randomly chosen in $S$, and $a$ is taken using the policy derived from $Q$. After observing $r$, the agent goes from state $s$ to $s'$ and the $Q$ function is updated following the Bellman equation \cite{bellman}:
\begin{equation}
\label{eq:bellman}
Q(s,a) \leftarrow Q(s,a) + \alpha [r + \gamma \cdot \max_{a'} Q(s',a') - Q(s,a)]
\end{equation}
where $\alpha$ is the learning rate that controls algorithm convergence and $\gamma$ is the discount factor. The algorithm proceeds until the episode ends, i.e a terminal state is reached. Convergence is reached by recursively updating values of $Q$ via temporal difference  incremental learning \cite{Sutton1988}.

\subsection{Deep Q network}
\label{dqn}
If the states are discrete, $Q$ function is represented as a table. However, when the number of states is too large, or the state space is continuous, this formulation becomes unfeasible. In such cases, the Q-function is computed as a parameterized non-linear function of both states and actions $Q(s,a;\theta)$ and the solution relies on finding the best parameters $\theta$. This can be learned by representing the Q-function using a DNN as shown in \cite{RLAvideogames} \cite{mnih2015humanlevel}, introducing deep Q-networks (DQN). 

The objective of a DQN is to minimize the mean square error (MSE) of the Q-values:
\begin{align}
\label{q-learn-loss}
L(\theta) = \frac{1}{2} [r + \max_{a'}Q(s',a'; \theta') - Q(s,a;\theta)]^2 \\
J(\theta) = \max_{\theta}[L(\theta)] 
\end{align}
\noindent Since this objective function is differentiable w.r.t $\theta$, the optimization problem can be solved using gradient based methods, e.g Stochastic Gradient Descent (SGD) \cite{Bottou2018OptimizationMF}.

\begin{figure*}[ht!]
	\centering
		\includegraphics[scale=1.0]{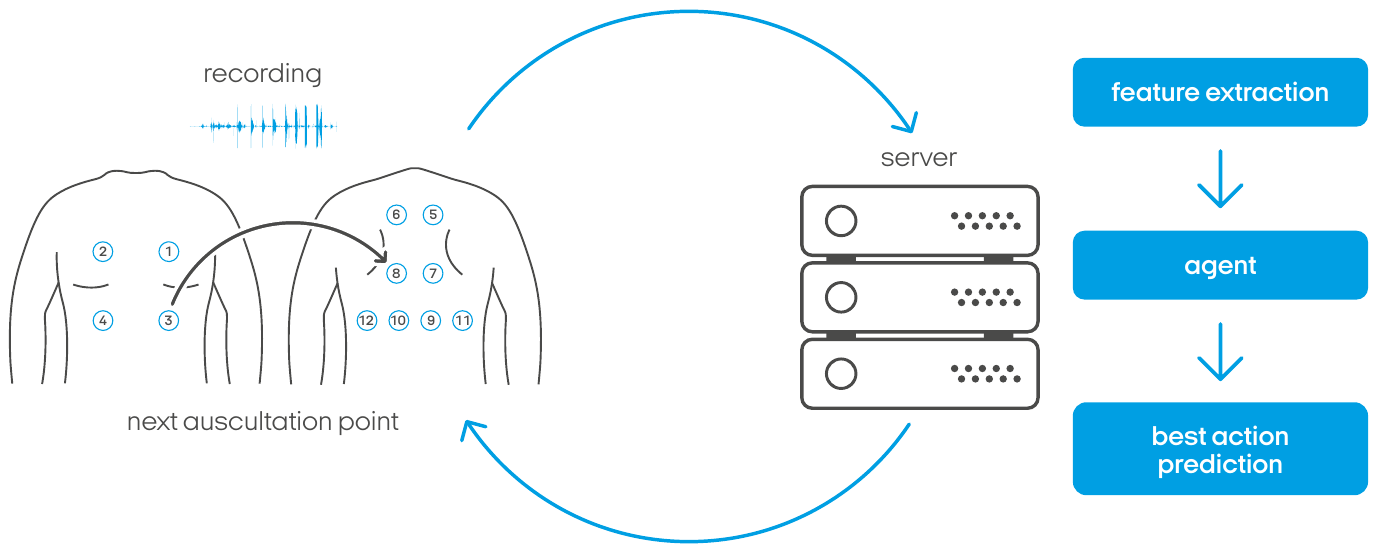}
	\caption{Interactive auscultation: the examiner starts auscultating the patient from the initial point (in this case point number 3), using our proprietary digital and wireless stethoscope, connected via Bluetooth to a smartphone. The recorded signal is sent to the server where a fixed set of features are extracted. These features represent the input to the agent that predicts the best action that should be taken. The prediction is then sent back to device and shown to the user, in this case to auscultate point number 8. The auscultation continues until agent is confident enough and declares predicted alarm value. The application works effectively even if the device is temporary offline: as soon as the connection is back, the agent can make decisions based on all the points that have been recorded so far.}
	\label{fig:system_overview}
\end{figure*}

\section{\uppercase{RL-based Interactive Auscultation}}
\label{sec:RL-based-interactive-auscultation}

\subsection{Problem Statement}
\label{problem-statmenent}
In our problem definition, the set of states $S$ is composed by the list of points already auscultated, each one described with a set of fixed number of features that characterize breath phenomena detected in that point. In other terms, $ S\in \mathbb{R}^{n \times m}$, where $n$ is the number of auscultation points and $m$ equals the number of extracted features per point, plus one for number of times this point has been auscultated.

The set of actions $A$, conversely, lie in a finite space: either auscultate another specified point (can be one of the points already auscultated), or predict diagnosis status of the patient if confident enough.

\subsection{Proposed Solution}
\label{proposed-solution}
With the objective of designing an agent that interacts with the environment described above, we adopted deep Q-learning as resolution algorithm. The proposed agent is a deep Q-network whose weights $\theta$ are updated following Eq. \ref{eq:bellman}, with the objective of maximizing the expected future rewards (Eq. \ref{q-learn-loss}):
\begin{equation}
\label{eq:updates}
\theta \leftarrow \theta + \alpha [r + \gamma \cdot \max_{a'} Q(s',a';\theta) - Q(s,a; \theta)] \nabla Q(s,a; \theta)
\end{equation}
where the gradient in Eq. \ref{eq:updates} is computed by backpropagation.
Similarly to what shown in \cite{RLAvideogames}, weight updates are performed through \textit{experience replay}. Experiences over many plays of the same game are accumulated in a replay memory and at each time step multiple Q-learning updates are performed based on experiences sampled uniformly at random from the replay memory. Q-network predictions map states to next action. Agent's decisions affect rewards signaling as well as the optimization problem of finding the best weights following Eq. \ref{eq:updates}.

The result of the auscultation of a given point is a feature vector of $m$ elements. After the auscultation, values from the vector are assigned to the appropriate row of the state matrix. Features used to encode agent's states are obtained after a feature extraction module whose core part consists of a convolutional recurrent neural network (CRNN) trained to predict breath phenomena events probabilities. The output of such network is a matrix whose rows show probability of breath phenomena changing over time. This data structure, called probability raster, is then post-processed in order to obtain $m$ features, representative of the agent's state.

Finally, reinforcement signals ($R$) are designed in the following way: rewards are given when the predicted diagnosis status is correct, penalties in the opposite case. Moreover, in order to discourage the agent of using too many points, a small penalty is provided for each additional auscultated point. The best policy for our problem is thus embodied in the best auscultation path, encoded as sequence of most informative APs to be analyzed, which should be as shortest as possible.

\subsection{Application}
\label{application}
The interactive auscultation application consists of two entities: the pair digital stethoscope and smartphone, used as the interface for the user to access the service; and a remote server, where the majority of the computation is done and where the agent itself resides. An abstraction of the entire system is depicted in Figure \ref{fig:system_overview}.

The first element in the pipeline is our proprietary stethoscope \cite{stethome}. It is a digital and wireless stethoscope similar to Littmann digital stethoscope \cite{littmann} in functionality, but equipped with more microphones that sample the signal at higher sampling rate, which enables it to gather even more information about the patient and background noise. The user interacts with the stethoscope through a mobile app installed on the smartphone, connected to the device via Bluetooth Low Energy protocol \cite{Gomez12overviewand}. Once the auscultation has started, a high quality recording of the auscultated point is stored on the phone and sent to the remote server. Here, the signal is processed and translated into a fixed number of features that will be used as input for the agent. The agent predicts which is the best action to perform next: it can be either to auscultate another point or, if the confidence level is high enough, return predicted patient's status and end the examination. Agent's decision is being made dynamically after each recording, based on breath phenomena detected so far. This allows the agent to make best decision given limited information which is crucial when the patient is an infant and auscultation gets increasingly difficult over time.

\begin{figure*}[ht!]
	\centering
		\includegraphics[scale=0.85]{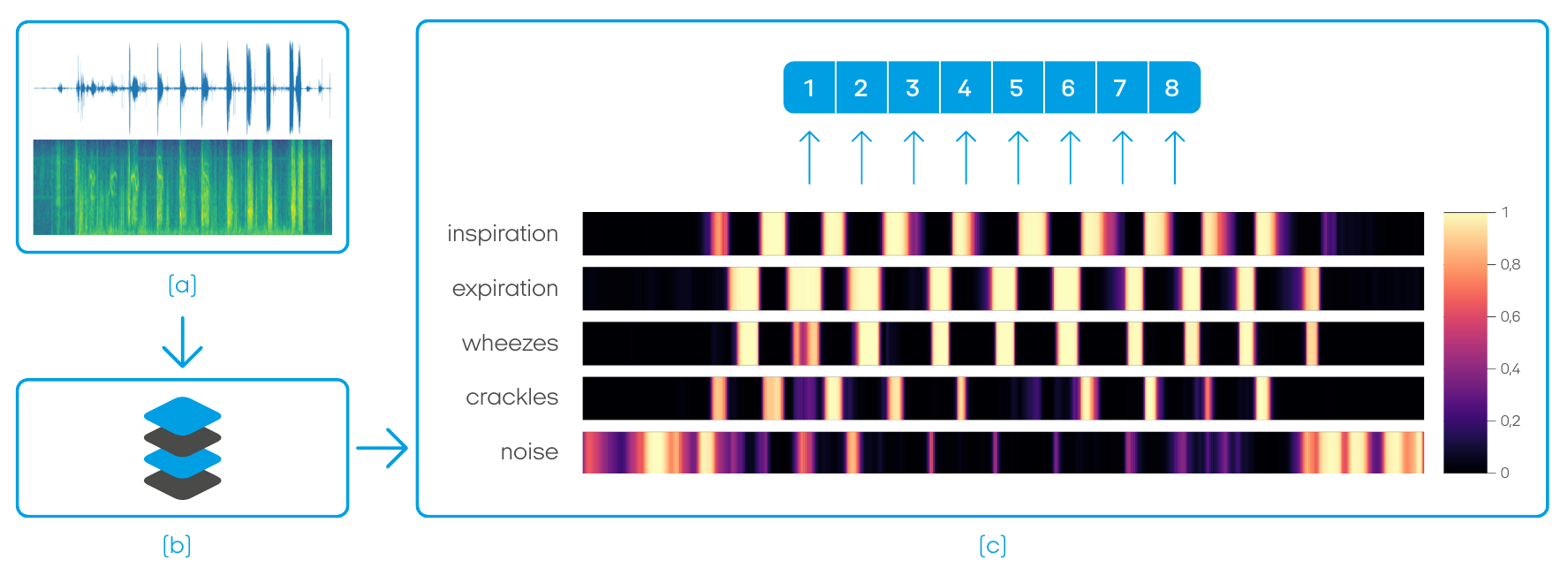}
	\caption{Feature extraction module: audio signal is first converted to spectrogram and subsequently fed to a CRNN, which outputs a prediction raster of $5$ classes: inspiration, expiration, wheezes, crackles and noise. This raster is then post-processed with the objective of extracting values representative of detection and intensity level of the critical phenomena, i.e wheezes and crackles. More specifically, maximum probability value and relative duration of tags are computed per each inspiration/expiration and the final features are computed as the average of these two statistics along all inspirations/expirations.}
	\label{fig:feature-extractor}
\end{figure*}
\section{\uppercase{Evaluation}}
\label{evaluation}
This section describes the experimental framework used to simulate the interactive auscultation application described in Subsection \ref{application}. In particular, in Subsection \ref{dataset} the dataset used for the experiments is described. In Subsection \ref{feature_extractor} a detailed description of the feature extraction module already introduced in Subsection \ref{proposed-solution} is provided. In Subsection \ref{RLA} the interactive agent itself is explained, while in Subsection \ref{exp} the final experimental setup is presented.

\subsection{Dataset}
\label{dataset}
Our dataset consists of a total of $570$ real examinations conducted in the Department of Pediatric Pulmonology of \textit{Karol Jonscher Clinical Hospital} in Pozna\'n (Poland) by collaborating doctors. Data collection involved young individuals of both genders ($46 \%$ females and $54 \%$ males.) and different ages: $13 \%$ of them were infants ($[0,1)$ years old), $40 \%$ belonging to pre-school age ($[1, 6)$) and $43 \%$ in the school age ($[6, 18)$).

Each examination is composed of $12$ APs, recorded in pre-determined locations (Figure \ref{fig:system_overview}). Three possible labels are present for each examination: $0$, when no pathological sounds at all are detected and there's no need to consult a doctor; $1$, when minor (innocent) auscultatory changes are found in the recordings and few pathological sounds in single AP are detected, but there's no need to consult a doctor; $2$, significant auscultatory changes, i.e major auscultatory changes are found in the recordings and patient should consult a doctor. This ground truth labels were provided by 1 to 3 doctors for each examination, in case there was more than one label the highest label value was taken. A resume of dataset statistics is shown in Table \ref{tab:dataset}. \\

\begin{table}[ht!]
\centering
\caption{Number of examinations for each of the classes.}
\begin{tabular}{llllr}
\cmidrule{1-3}
Label & Description & $N_{examinations}$\\
\midrule
$0$ & no auscultatory \\ & changes - \textit{no alarm} & $200$\\
$1$ & innocent auscultatory \\ & changes - \textit{no alarm} & $85$  \\
$2$ & significant auscultatory \\ & changes - \textit{alarm} & $285$  \\
\bottomrule
\end{tabular}
\label{tab:dataset}
\end{table}

\subsection{Feature Extractor}
\label{feature_extractor}

The features for the agent are extracted by a feature extractor module that is composed of three main stages, schematically depicted in Figure \ref{fig:feature-extractor}: at the beginning of the pipeline, the audio wave is converted to its magnitude spectrogram, a representation of the signal that can be generated by applying short time fourier transform (STFT) to the signal (a). The time-frequency representation of the data is fed to a convolutional recurrent neural network (b) that predicts breath phenomena events probabilities in form of predictions raster. Raster is finally post-processed in order to extract $8$ interesting features (c).

\subsubsection{CRNN}
This neural network is a modified implementation of the one proposed by {\c{C}}akir et al. (2017), i.e a CRNN designed for polyphonic sound event detection (SED). In this structure originally proposed by  the convolutional layers act as pattern extractors, the recurrent layers integrate the extracted patterns over time thus providing the context information, and finally the feedforward layer produce the activity probabilities for each class \cite{DBLP:journals/corr/CakirPHHV17}. We decided to extend this implementation including dynamic routing \cite{Sabour2017DynamicRB} and applying some key ideas of Capsule Networks (CapsNet), as suggested in recent advanced studies \cite{capsulenetworkspsed} \cite{capsulenetworkspsed2}. The CRNN is trained to detect $5$ types of sound events, namely: inspirations, expirations, wheezes, crackles \cite{auscultation} and noise. \\

\begin{figure*}[ht!]
\centering
  \subfigure[]{\includegraphics[scale=0.45]{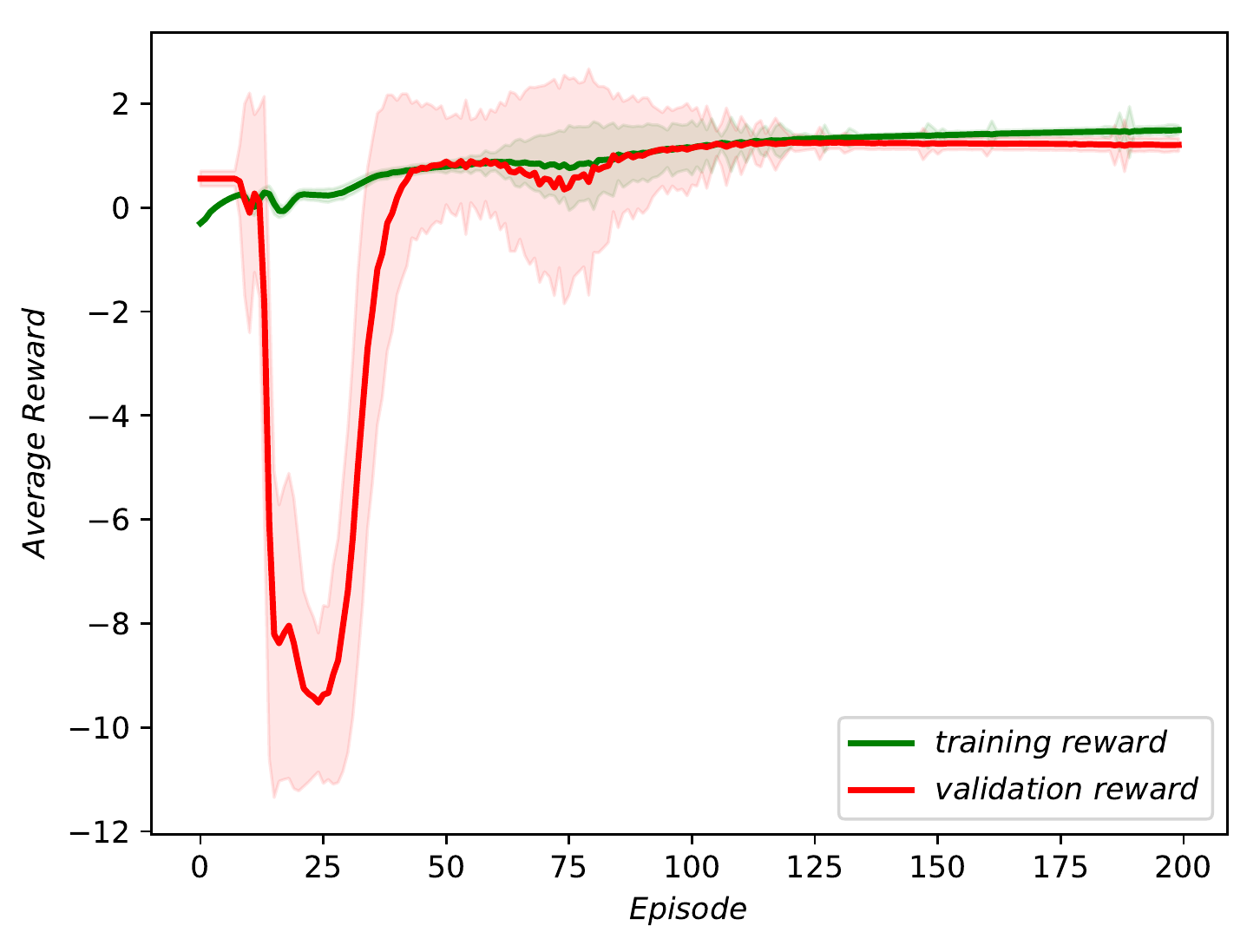}}\quad
  \subfigure[]{\includegraphics[scale=0.45]{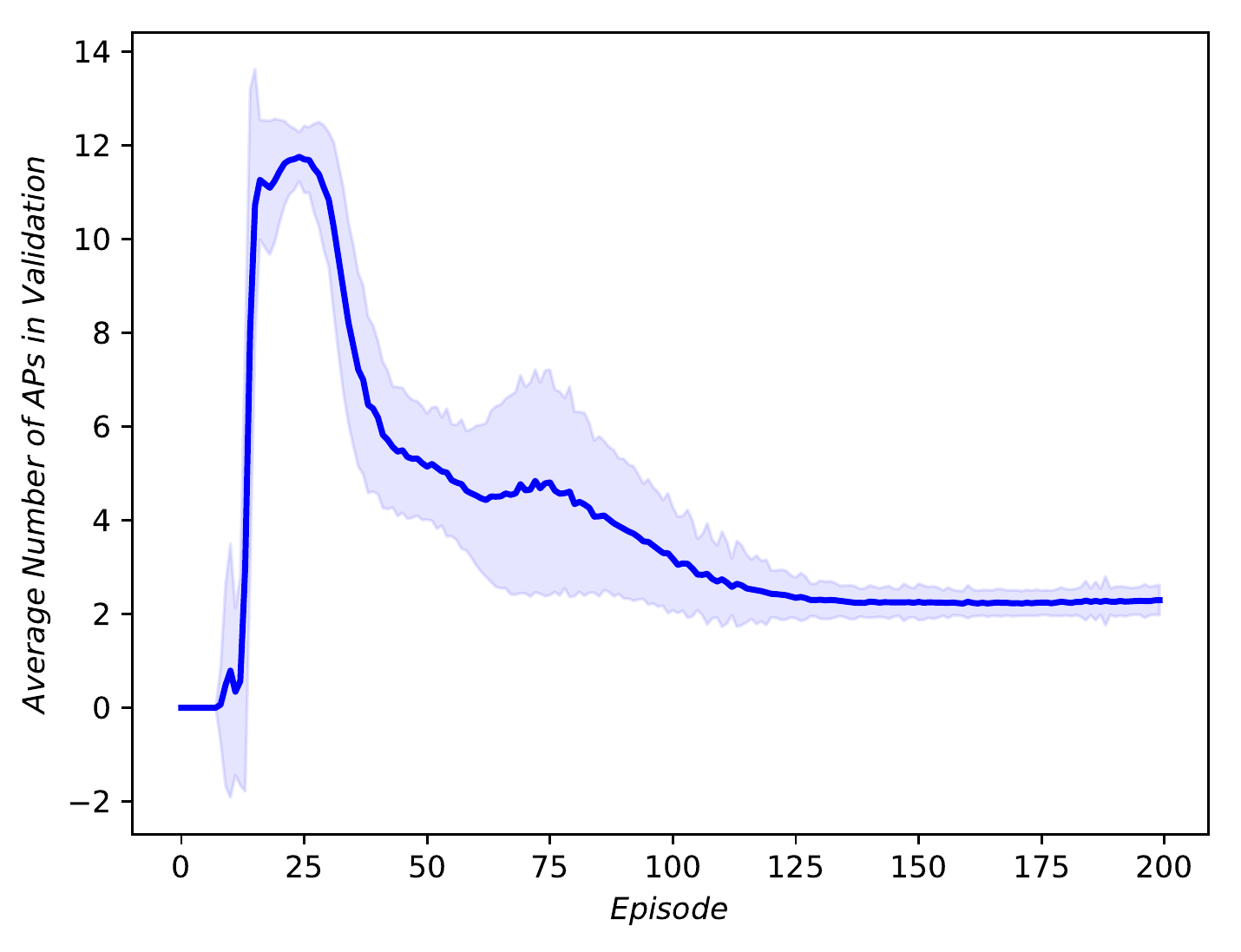}}
\caption{Interactive agent learning curves: in the very first episodes the agent randomly guesses the state of the patient, without auscultating any point. Next comes the exploration phase when the agent auscultates many points, often reaching the 12-point limit which results in high penalties. Finally, as the agent plays more and more episodes, it starts learning the optimal policy using fewer and fewer points until he finds the optimal solution}
  \label{learning-curves}
\end{figure*}

\subsubsection{Raster Post-processing}
Wheezes and crackles are the two main classes of pathological lung sounds. The purpose of raster post-processing is to extract a compact representation that will be a good descriptions of their presence/absence and level of intensity. Thus, for each inspiration and expiration event we calculate two values for each pathological phenomena (wheezes and crackles): maximum probability within said inspiration/expiration and relative duration after thresholding (the level in which the inspiration/expiration is filled, or covered with the pathological phenomenon). All extracted values are then averaged across all inspirations and expirations separately. We therefore obtain $8$ features: average maximum wheeze probability on inspirations (1) and expirations (2), average relative wheeze length in inspirations (3) and expirations (4) and the same four features (5, 6, 7 and 8) for crackles.

\begin{figure*}[ht!]
	\centering
		\includegraphics[scale=1.01]{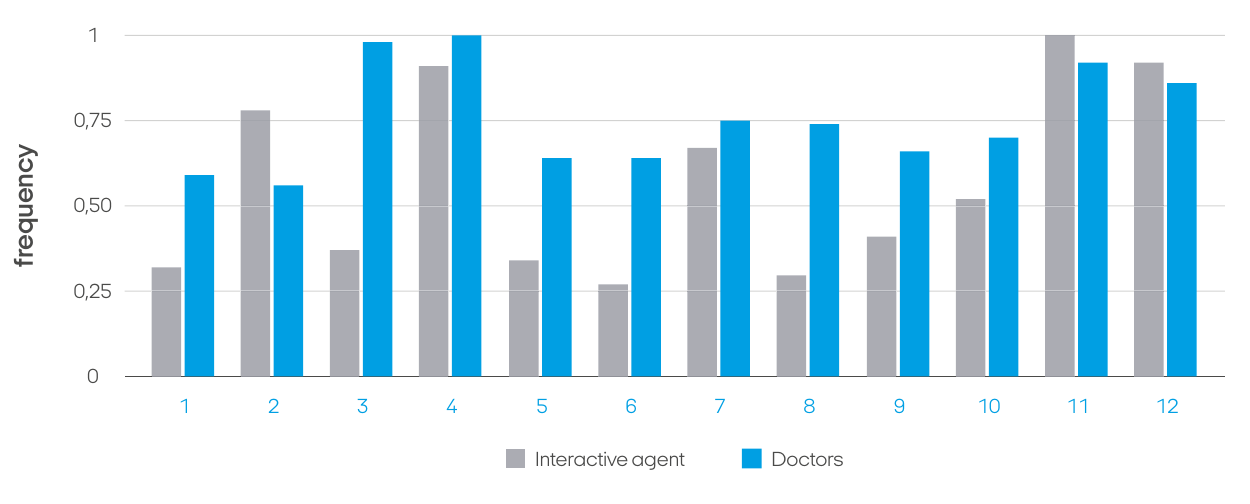}
	\caption{Histograms: we compared the distribution of most used points by the agent against the ones that doctors would most often choose at examination time. Results show that the agent learned importance of points  without any prior knowledge about human physiology}
	\label{fig:histograms}
\end{figure*}

\subsection{Reinforcement Learning Agent}
\label{RLA}
Our agent consists of a deep fully connected neural network. The network takes as input a state matrix of size $12$ rows $\times \ 9$ columns; then processes the input through $3$ hidden layers with $256$ units each followed by ReLU nonlinearity \cite{Nair:2010:RLU:3104322.3104425}; the output layer is composed by $15$ neurons which represent expected rewards for each of the possible future actions. This can be either to request one of the $12$ points to be auscultated, or declare one of the three alarm status, i.e predict one of the three labels and finish the examination.

The state matrix is initially set to all zeros, and the $i_{th}$ row is updated each time $i_{th}$ AP is auscultated. First $8$ columns of state matrix correspond to eight features described in previous section, while the last value is a counter for the number of times this auscultation point was auscultated. At every interaction with the environment, the next action $a$ to be taken is defined as $arg \max$ of the output vector. At the beginning of the agent's training we ignore agent's preferences and perform random actions, as the training proceeds we start to use agent's recommended actions more and more often. For a fully trained model, we always follow agent's instructions. The agent is trained to predict three classes, but classes $0$ and $1$, treated as agglomerated \textit{not alarm} class, are eventually merged at evaluation phase.

The agent is trained with objective of minimizing Eq. \ref{q-learn-loss}, and Q-values are recursively updated by temporal difference incremental learning. There are two ways to terminate the episode: either make a classification decision, getting the reward/penalty that follows table \ref{rla-rewards-table}; or reaching a limit of $12$ actions which results in a huge penalty of $r = -10.0$. Moreover, when playing the game a small penalty of $r = -0.01$ is given for each requested auscultation, this is to encourage the agent to end the examination if it doesn't expect any more information coming from continued auscultation.

\begin{table}[ht]
\caption{Reward matrix for Reinforcement Learning Agent final decisions.}
\label{rla-rewards-table}
\centering
\begin{tabular}{cllll}
&                        & \multicolumn{3}{l}{predicted}                                            \\ \cline{3-5} 
& \multicolumn{1}{l|}{}  & \multicolumn{1}{l|}{0} & \multicolumn{1}{l|}{1} & \multicolumn{1}{l|}{2} \\ \cline{2-5} 
\multicolumn{1}{c|}{\multirow{3}{*}{\rotatebox{90}{actual}}} & \multicolumn{1}{l|}{0} & \multicolumn{1}{l|}{2.0}  & \multicolumn{1}{l|}{0.0}  & \multicolumn{1}{l|}{-1.0}  \\ \cline{2-5} 
\multicolumn{1}{c|}{}                        & \multicolumn{1}{l|}{1} & \multicolumn{1}{l|}{0.0}  & \multicolumn{1}{l|}{2.0}  & \multicolumn{1}{l|}{-0.5}  \\ \cline{2-5} 
\multicolumn{1}{c|}{}                        & \multicolumn{1}{l|}{2} & \multicolumn{1}{l|}{-1.0}  & \multicolumn{1}{l|}{-0.5}  & \multicolumn{1}{l|}{2.0}  \\ \cline{2-5} 
\end{tabular}
\end{table}

\subsection{Experimental Setup}
\label{exp}
We compared the performance of reinforcement learning agent, from now on referred to as \textit{interactive} agent, to its \textit{static} counterpart, i.e an agent that always performs an exhaustive auscultation (uses all $12$ APs).
In order to compare the two agents we performed $5$-fold cross validation for $30$ different random splits of the dataset into training ($365$ auscultations), validation ($91$) and test ($114$) set. We trained the agent for $200$ episodes, setting $\gamma=0.93$ and using Adam optimization algorithm \cite{Kingma2014AdamAM} to solve Eq. \ref{q-learn-loss}, with learning rate initially set to $0.0001$. 

Both in validation and test phase, $0$ and $1$ labels were merged as single $not \ alarm$ classes. Therefore results shown in the following refer to the binary problem of alarm detection: we chose as comparative metrics balanced accuracy (BAC) defined as unweighted mean of sensitivity and specificity; and F1-score, harmonic mean of precision and recall, computed for each of the two classes. 

\section{\uppercase{Results}}
\label{results}
In Table \ref{results-experiments} we show the results of the experiments we conducted. The interactive agent performs the auscultation using on average only $3$ APs, effectively reducing the time of the examination $4$ times. This is a very significant improvement and it comes at a relatively small cost of 2.5 percent point drop in classification accuracy.

\begin{table}[ht!]
\caption{Results of experiments.}
\centering
\begin{tabular}{llllrr}
\cmidrule{1-5}
Agent & $BAC$ & $F1_{alarm}$  & $F1_{not \ alarm}$ & $APs$ \\
\midrule
Static & $84.8 \ \% $ & $82.6 \ \%$ & $85.1 \ \%$ & $12$ \\
Interactive & $82.3 \ \%$ & $81.8 \ \%$ & $82.6 \ \%$ & $3.2$ \\
\bottomrule
\end{tabular}
\label{results-experiments}
\end{table}

Figure \ref{learning-curves} shows learning curves of rewards and number of points auscultated by the agent. In the very first episodes the agent directly guesses the state of the patient, declaring the alarm value without auscultating any point. As soon as it starts exploring other possible action-state scenarios, it often reached the predefined limit of $12$ auscultation points which significantly reduces its average received reward. However, as it plays more episodes, it starts converging to the optimal policy, using less points on average.

In order to assess the knowledge learned by the agent, we conducted a survey involving a total of $391$ international experts. The survey was distributed among the academic medical community and in hospitals. In the survey we asked each participant to respond a number of questions regarding education, specialization started or held, assessment of their own skills in adult and child auscultation, etc. In particular, we asked them which points among the $12$ proposed would be auscultated more often during an examination. Results of the survey are visible in Figure \ref{fig:histograms} where we compare collected answers with most used APs by the interactive agent. It's clear that the agent was able to identify which APs carry the most information and are the most representative to the overall patient's health status. This is the knowledge that all human experts gain from many years of clinical practice. In particular the agent identified points 11 and 12 as very important. This finding is confirmed by the doctors who strongly agree that these are the two most important APs on patient's back. On the chest both doctors and the agent often auscultate point number 4, but the agent prefers point number 2 instead of 3, probably due to the distance from the heart which is a major source of interference in audio signal.

The agent seems to follow two general rules during the auscultation: firstly, it auscultates points belonging both to the chest and to the back; secondly, it tries to cover as much area as possible, visiting not-subsequent points. For instance, the top $5$ auscultation paths among the most repeating sequences that we observed are: $[4,9,11]$, $[8,2,9]$, $[2,11,12]$, $[7,2,8]$, $[4,11,12]$. These paths cover only $3 \%$ of the possible paths followed by the agent: this means the agent does not follow a single optimal path or even couple of paths, but instead uses a wide variety of paths depending on breath phenomena detected during the examination.

\section{\uppercase{Conclusions}}
\label{conclusions}

We have presented a unique application of reinforcement learning for lung sounds auscultation, with the objective of designing an agent being able to perform the procedure interactively in the shortest time possible. 

Our interactive agent is able to perform an intelligent selection of auscultation points. It performs the auscultation using only $3$ points out of a total of $12$, reducing fourfold the examination time. In addition to this, no significant decrease in diagnosis accuracy is observed, since the interactive agent gets only $2.5$ percent points lower accuracy than its static counterpart that performs an exhaustive auscultation using all available points.

Considering the research we have conducted, we believe that further improvements can be done in the solution proposed. In the near future, we would like to extend this work to show that the interactive solution can completely outperform any static approach to the problem. We believe that this can be achieved by increasing the size of the dataset or by more advanced algorithmic solutions, whose investigation and implementation was out of the scope of this publication. 

\section*{\uppercase{Copyright Notice}}
This contribution has been published in the proceedings of the 11\textsuperscript{th} International Conference on Agents and Artificial Intelligence (ICAART) - Volume 1. Conference link: http://www.icaart.org/?y=2019

\bibliographystyle{apalike}
{\small
\bibliography{references}}

\end{document}